\documentclass[conference]{IEEEtran}
\ifCLASSINFOpdf
\else
\fi
\usepackage{amsmath,dsfont,bbm,epsfig,amssymb,amsfonts,amstext,verbatim,amsopn,cite,subfigure,multirow,multicol,lipsum,xfrac}
\usepackage{amsthm}
\usepackage{mathtools,amsthm}
\usepackage{perpage}
\usepackage{balance}
\usepackage{url}
\usepackage{amsfonts}
\usepackage{epsfig}
\usepackage[font={small}]{caption}
\usepackage{psfrag}	
\usepackage{etoolbox}
\usepackage{algorithmicx}
\usepackage[Algorithm,ruled]{algorithm}
\usepackage{algpseudocode}
\usepackage{pifont}
\usepackage[utf8]{inputenc}
\usepackage[T1]{fontenc}  
\usepackage[nolist]{acronym}
\MakePerPage{footnote}
\usepackage{paralist}
\usepackage{enumitem}
\usepackage{bbm}
\usepackage[process=auto]{pstool}
\usepackage{tikz}
\usetikzlibrary{shapes,arrows}
\usepackage{color}
\definecolor{myblue}{rgb}{0.07, 0.41, 0.7}
\definecolor{myred}{rgb}{0.85, 0.33, 0.1}
\newcommand{\setX}{\mathbbmss{X}}

\newcommand{\setR}{\mathbbmss{R}}

\newcommand{\setB}{\mathbbmss{B}}

\newcommand{\setC}{\mathbbmss{C}}

\newcommand{\setN}{\mathbbmss{N}}

\newcommand{\sgn}[1]{\mathrm{sgn}\left( #1 \right)}

\DeclareMathOperator*{\argmin}{\mathrm{argmin}\hspace*{1mm}}
\DeclareMathOperator*{\argmax}{\mathrm{argmax}\hspace*{1mm}}

\newcommand{\rmp}{\mathrm{p}}

\newcommand{\rmR}{\mathrm{R}}

\newcommand{\rmG}{\mathrm{G}}
\newcommand{\rmQ}{\mathrm{Q}}

\newcommand{\her}{\mathsf{H}}

\newcommand{\sfd}{\mathsf{d}}

\newcommand{\sfp}{\mathsf{p}}

\newcommand{\man}{\mathcal{N}}

\newcommand{\bx}{{\boldsymbol{x}}}
\newcommand{\hx}{{\hat{x}}}

\newcommand{\vv}{\mathrm{v}}
\newcommand{\hxx}{\hat{\mathrm{x}}}
\newcommand{\xx}{\mathrm{x}}
\newcommand{\yy}{\mathrm{y}}
\newcommand{\zz}{\mathrm{z}}

\newcommand{\bhx}{{\boldsymbol{\hat{x}}}}
\newcommand{\set}[1]{\left\lbrace#1\right\rbrace}

\newcommand{\bz}{{\boldsymbol{z}}}

\newcommand{\bv}{{\boldsymbol{v}}}

\newcommand{\bgg}{{\mathbf{g}}}

\newcommand{\dif}{\mathrm{d}}

\newcommand{\by}{{\boldsymbol{y}}}
\newcommand{\bt}{{\boldsymbol{t}}}

\newcommand{\mA}{\mathbf{A}}

\newcommand{\ma}{\mathbf{A}}
\newcommand{\mI}{\mathbf{I}}
\newcommand{\mone}{\mathbf{1}}
\newcommand{\mJ}{\mathbf{J}}

\newcommand{\mU}{\mathbf{U}}
\newcommand{\mD}{\mathbf{D}}

\newcommand{\md}{\mathrm{D}}

\newcommand{\E}[1]{\mathbb{E} \left\lbrace #1 \right\rbrace}

\newcommand{\norm}[1]{\lVert #1 \rVert}

\newcommand{\abs}[1]{\lvert #1 \rvert}

\newtheoremstyle{mystyle}
  {}
  {}
  {}
  {}
  {\bfseries}
  {:}
  { }
  {}

\theoremstyle{mystyle}

\newtheorem{remark}{Remark}

\newcounter{bar}

\begin{document}
\title{RLS Recovery with Asymmetric Penalty: Fundamental Limits and Algorithmic Approaches}
\author{
\IEEEauthorblockN{
Ali Bereyhi\IEEEauthorrefmark{1},
Mohammad Ali Sedaghat\IEEEauthorrefmark{2},
Ralf R. M\"uller\IEEEauthorrefmark{1},
}
\IEEEauthorblockA{
\IEEEauthorrefmark{1}Institute for Digital Communications (IDC), Friedrich-Alexander Universit\"at Erlangen-N\"urnberg\\
\IEEEauthorrefmark{2}Cisco Optical GmbH, N\"urnberg\\
ali.bereyhi@fau.de, 
msedaghat@cisco.com, 
ralf.r.mueller@fau.de
\thanks{This work was supported by the German Research Foundation, Deutsche Forschungsgemeinschaft (DFG), under Grant No. MU 3735/2-1.}
}
}

\IEEEoverridecommandlockouts

\maketitle
\tikzstyle{block} = [draw, rounded corners, rectangle, minimum height=2.5em, minimum width=5em]
\tikzstyle{margin} = [draw, dotted, rectangle, minimum height=2.1em, minimum width=2em]
\tikzstyle{sum} = [draw, circle, node distance=1cm, inner sep=0pt]
\tikzstyle{input} = [coordinate]
\tikzstyle{output} = [coordinate]
\tikzstyle{pinstyle} = [pin edge={to-,thick,black}]

\begin{acronym}
\acro{ls}[LS]{Least Squares}
\acro{rls}[RLS]{Regularized LS}
\acro{cs}[CS]{Compressive Sensing}
\acro{mimo}[MIMO]{Multiple-Input Multiple-Output}
\acro{csi}[CSI]{Channel State Information}
\acro{awgn}[AWGN]{Additive White Gaussian Noise}
\acro{iid}[i.i.d.]{independent and identically distributed}
\acro{ut}[UT]{User Terminal}
\acro{bs}[BS]{Base Station}
\acro{tas}[TAS]{Transmit Antenna Selection}
\acro{lse}[LSE]{Least Square Error}
\acro{rhs}[r.h.s.]{right hand side}
\acro{lhs}[l.h.s.]{left hand side}
\acro{wrt}[w.r.t.]{with respect to}
\acro{rs}[RS]{Replica Symmetry}
\acro{rsb}[RSB]{Replica Symmetry Breaking}
\acro{papr}[PAPR]{Peak-to-Average Power Ratio}
\acro{rzf}[RZF]{Regularized Zero Forcing}
\acro{snr}[SNR]{Signal-to-Noise Ratio}
\acro{mse}[MSE]{Mean Square Error}
\acro{amp}[AMP]{Approximate Message Passing}
\acro{gamp}[GAMP]{Generalized AMP}
\acro{vamp}[VAMP]{Vector AMP}
\end{acronym}

\begin{abstract}
This paper studies regularized least square recovery of signals whose samples' prior distributions are nonidentical, e.g., signals with time-variant sparsity. For~this model, Bayesian framework suggests to regularize the least squares term with~an asymmetric penalty. We investigate this problem in two respects: First, we characterize the asymptotic performance via the replica method and then discuss algorithmic approaches to the problem. Invoking the asymptotic characterization of the performance, we propose a tuning strategy to optimally tune the algorithmic approaches for recovery. To demonstrate applications of~the~results, the particular example of BPSK recovery is investigated and the efficiency of the proposed strategy is depicted in the shadow of results available in the literature.\vspace*{3mm}
%
\end{abstract}
\section{Introduction}
\label{sec:intro}
Reconstructing a signal from a set of noisy, and generally underdetermined, observations is a classical problem in statistics. In this problem, the vector of signal samples $\bx\in \setX^N$ with $\setX\subset \setC$ is linearly measured via the measuring matrix $\mA\in \setC^{M\times N}$ and observed as
\begin{align}
\by = \mA \bx + \bz \label{eq:system1}
\end{align}
where $\bz\sim \mathcal{CN}(0,\sigma^2 \mI_M)$ denotes measuring noise.~The~classical recovery approach is to find the \ac{ls}~solution which minimizes the error term $\norm{\by-\ma \bv}^2$ over $\bv\in\setX^N$. For a large scope of applications, it is known in advance that the signal fulfills some properties, e.g., sparsity. In this case, Bayesian inference suggests to modify \ac{ls} recovery as
\begin{align}
\bhx = \bgg(\by|\mA) = \argmin_{\bv\in\setX^N} \frac{1}{\lambda}\norm{\by-\ma \bv}^2 + u_\vv(\bv)
\label{eq:LSE}
\end{align}
for some $\lambda\in\setR^+$. Here, non-negative penalty $u_\vv(\cdot)$ regularizes the \ac{ls} recovery with respect to the prior information on the signal. The estimator in \eqref{eq:LSE} encloses a large class of regression methods employed in signal processing and machine learning. A well-known application of this \ac{rls} estimator is for sparse recovery in \ac{cs} \cite{donoho2006compressed,candes2006robust}. In \ac{cs}, the number of measurements $M$ is smaller than the number of samples $N$, and the signal is known~to~be~sparse. The formulation in \eqref{eq:LSE} addresses several recovery algorithms in \ac{cs}. An example is LASSO in which $u_\vv(\bv)=\norm{\bv}_1$.

Regularization in \ac{rls} recovery is conventionally assumed to be ``symmetric''. For example in LASSO,~all~entries of~$\bv$ appear in the regularization term similarly. In the Bayesian framework, symmetric regularization means that the estimator postulates the signal samples are generated \ac{iid}. In other words, it is assumed that the \textit{temporal} correlation among the samples is negligible and the signal exhibits identical variation over time. Considering proper sampling and preprocessing techniques, the~former~assumption often holds in practice. However, there are several scenarios in which the latter~assumption~is~not~realistic,~e.g., sparse signals with time-variant sparsity~\cite{khajehnejad2011analyzing,vaswani2010modified,jacques2010short,scarlett2013compressed}.~In~these~scenarios, \ac{rls} recovery can be considered with ``asymmetric'' penalty; see \cite{rauhut2016interpolation, bah2016sample,oymak2012recovery,mansour2012support, khajehnejad2009weighted,candes2008enhancing,bereyhi2018maximum} for particular examples in \ac{cs}.

The recovery scheme in \eqref{eq:LSE} is iteratively implemented~with linear complexity via a standard \ac{amp} algorithm \cite{donoho2009message,donoho2010message,rangan2011generalized,rangan2017vector}. The performance of the algorithm, however, depends on the choice of free parameters, e.g., $\lambda$, in the original \ac{rls} problem. In this respect, a systematic strategy is required in practice to tune these free parameters, such that the performance of the algorithm is optimized.

\subsection*{Contributions}
In \cite{bereyhi2018maximum}, the asymptotic performance of \ac{rls} recovery~with asymmetric regularization was derived for real-valued setups. This paper completes the study in \cite{bereyhi2018maximum} by investigating~the algorithmic approaches to asymmetric \ac{rls} recovery.~The~main contributions of this study are 
\begin{inparaenum}
\item[($i$)] Extending the large-system results of \cite{bereyhi2018maximum} to complex-valued settings, and
\item[($ii$)] Proposing a novel tuning strategy for an algorithmic approach.
\end{inparaenum}

To illustrate an application of the results, we consider the problem of BPSK recovery studied formerly in \cite{thrampoulidis2015regularized,atitallah2017ber}. It is shown that the tuning results in \cite{atitallah2017ber} can be straightforwardly recovered from our proposed tuning strategy.

%
\section{Problem Formulation}
We consider the setting in \eqref{eq:system1} in which the measuring matrix $\mA \in \setC^{M\times N}$ is generated randomly. It is moreover assumed that $\mJ=\mA^{\her} \mA$ has the decomposition $\mJ= \mU \mD \mU^{\her}$ where $\mU$ is a Haar distributed random matrix\footnote{This means that $\mU$ is uniformly distributed over the unitary group} and $\mD$ denotes the diagonal matrix of eigenvalues. We denote the asymptotic empirical distribution of the eigenvalues of $\mJ$ by $\mathrm{p}_{\mJ}(\lambda)$. 

We consider a set of \ac{rls} recovery problems in which the number of measurements $M$ is a deterministic sequence of the number of signal samples $N$ such that the compression ratio $\rho \coloneqq {M}/{N}$ is kept fixed and bounded as $N$ grows large.

\subsection{Signal Model}
\label{sec:signal_model}
The signal samples are considered to be independent with nonidentical prior distributions. To illustrate the signal model precisely, let the set of sample indices $[N]\coloneqq\set{1,\ldots.N}$ be partitioned into disjoint subsets $\setN_j$ for $j\in[J]$ where $J$ is bounded. The samples in $\bx$ are then partitioned into $J$ blocks as follows: The $j$-th block $\setB_j (\bx)$ contains samples whose~in-dices are in $\setN_j$. This means that
\begin{align}
\setB_j (\bx) \coloneqq \set{x_n:  n\in\setN_j }. \label{eq:blockAAAAAAA}
\end{align}
The definition in \eqref{eq:blockAAAAAAA} implies that $\bx \equiv \cup_{j=1}^J \setB_j(\bx)$. 

For $n\in[N]$, the entry $x_n$ is assumed to be distributed with $\rmp_{j(n)}(x_n;\mu_{n})$ where $\rmp_{j}(x;\mu)$ for $j\in[J]$ is a probability distribution with the ``control factor'' $\mu$ and $j(n)$ denotes the index of the block which encloses $x_n$. 
\subsubsection*{Example} 
An example of this model is a signal with~multiple priors whose sparsity varies over time: Consider a sparse signal consisting of two equal-sized parts. Each part represents a statistically different phenomenon. The non-zero samples of the first and second parts are distributed with $\rmp_1(x)$ and $\rmp_2(x)$, respectively. Moreover, the signal has a non-uniform sparsity pattern, such the sparsity factors\footnote{By the sparsity factor, we mean the fraction of entries which are non-zero.} of the first and second part are the distinct numbers $\mu_1$ and $\mu_2$, respectively. For this~sig-nal, the samples are divided into $J=2$ blocks with
\begin{align}
\setN_j = \set{(j-1)N/2+1, \ldots,jN/2}.
\end{align}
for $j=1,2$. The sample $x_n$ in block $j$ is distributed with 
\begin{align*}
\rmp_j(x_n;\mu_n)=\mu_n \rmp_j(x_n) + (1-\mu_n) \delta(x_n)
\end{align*}
where 
\begin{align}
\mu_n=\begin{cases}
\mu_1 & n\in\setN_1\\
\mu_2 & n\in\setN_2
\end{cases}
\end{align}
For $\rmp_1(x)=\rmp_2(x)$, the model reduces to a sparse signal with a non-uniform sparsity pattern.

\subsection{Asymmetric \ac{rls} Recovery}
To recover $\bx$ from the noisy observation $\by$, we employ the \ac{rls} algorithm in \eqref{eq:LSE}. Noting that the signal has a non-uniform prior, we consider following generic asymmetric penalty
\begin{align}
u_\vv(\bv) = \sum_{j=1}^J \sum_{n\in\setN_j} u_{j}(v_n;w_n)= \sum_{n=1}^N u_{j(n)}(v_n;w_n) \label{eq:penalty}
\end{align}
where $u_{j}(v;w)$ for $j\in[J]$ is a non-negative penalty function tuned by some scalar $w$. $\set{w_n}$ is moreover a sequence of tunable factors which are tuned such that the performance of the recovery algorithm in optimized. A classical example of such penalty is the regularization term in the weighted $\ell_1$-norm \ac{rls} scheme; see for example \cite{bah2016sample}. In this example, $J=1$ and we have $u(v_n;w_n)=w_n\abs{v_n}$.

To implement the \ac{rls} algorithm in \eqref{eq:LSE} with asymmetric penalty \eqref{eq:penalty}, we need to solve an optimization problem. For some choices of $u_\vv(\bv)$ and $\setX$, this optimization can be posed as a linear programming or iteratively solved via \ac{amp} \cite{donoho2009message,donoho2010message,rangan2011generalized}. There are moreover various regularization terms and signal supports for which the \ac{rls} algorithm is computationally infeasible to implement. 
We therefore intend to answer the following questions in this paper:
\begin{inparaenum}
\item[(a)] For given penalty, what is the asymptotic input-output distortion defined as 
\begin{align}
D = \frac{1}{N} \E{\sum_{n=1}^N \sfd(\hx_n;x_n)} \label{eq:dist}
\end{align}
for a desired distortion function $\sfd(\cdot;\cdot)$ when $N$ grows large? 
\item[(b)] When \ac{rls} recovery is implemented via an iterative~algorithm, what is the optimal strategy to set the tunable factors, i.e., $\set{w_n}$ and $\lambda$, such that the asymptotic distortion~$D$~is~minimized?
\end{inparaenum}

The first question is answered via the replica method which has been widely employed in this context; see \cite{tanaka2010optimal,rangan2012asymptotic,tulino2013support,vehkapera2014analysis,bereyhi2016statistical,bereyhi2018maximum} and references therein. To address the latter question, we utilize our asymptotic results to propose a tuning strategy. 
%
%
%
\section{Asymptotics via the Replica Method}
For real-valued settings, i.e., when $\setX\subset \setR$ and $\mA\in\setR^{M\times N}$, the asymptotic distortion $D$ has been derived in \cite{bereyhi2018maximum} via the replica method. In this section, we extend the results to complex-valued setups. The validity of the derivations follow several conjectures which we skip the details for simplicity. Assuming these conjectures to hold, the asymptotic distortion is given in a simple form depending on the so-called ``replica symmetry'' property of the problem. 
From the literature, it is known that  \ac{rls} recovery is replica symmetric for a large class of penalties \cite{bereyhi2016statistical,reeves2016replica}. For cases in which the problem does not exhibit replica symmetry, the asymptotic distortion is derived by extending the analyses in \cite{bereyhi2018maximum} and following the approach~in~\cite{bereyhi2016statistical}.
\subsection{Asymptotic Characterization}
As $N$ grows large, \ac{rls} recovery exhibits the decoupling property which lets us characterize the large-system properties of the setting in a simple form. This property indicates that statistics of any pair $(x_n, \hx_n)$ can be described via an equivalent scalar system which models the impact of interference from other samples via effective noise. Such a property has been studied for several recovery schemes in various settings \cite{guo2005randomly,rangan2012asymptotic,bereyhi2016rsb,bereyhi2016statistical}. In this section, we represent the decoupling property of \ac{rls} recovery in \eqref{eq:LSE} with penalty \eqref{eq:penalty}. For this aim, let us first define the ``decoupled'' system.

The decoupled system consists of the sample $\xx_n$ distributed with $\rmp_{j(n)}(\xx_n;\mu_n)$, the noisy observation $\yy_n=\xx_n+\zz$ with $\zz\sim\mathcal{CN}(0,\theta^2)$ and the \ac{rls} recovery algorithm 
\begin{align}
\hxx_n = \argmin_{v\in\setX} \frac{1}{\tau}\abs{\yy_n- v}^2 + u_{j(n)}(v;w_n)
\label{eq:decop}
\end{align}
for some tunable factor $\tau$. The noise variance $\theta^2$ and the factor $\tau$ in the decoupled system are written in terms of some real scalars $\chi$ and $\sfp$ as $\tau = \left[ \rmR_\mJ(-\frac{\chi}{\lambda})\right]^{-1} {\lambda}$ and
\begin{align}
\theta^2 &= \left[ \rmR_\mJ(-\frac{\chi}{\lambda})\right]^{-2} \frac{\partial}{\partial\chi} \left[(\sigma^2 \chi-\lambda \sfp)\rmR_\mJ(-\frac{\chi}{\lambda})\right] 
\end{align}
where $\rmR_\mJ (\cdot)$ denotes the $\rmR$-transform of $\rmp_\mJ (\lambda)$ and is defined as $\rmR_\mJ (\omega) \coloneqq \rmG_\mJ^{-1} (-\omega) - \omega^{-1}$ with $\rmG_\mJ^{-1} (\cdot)$ being the inverse with respect to composition of 
$\rmG_{\mJ}(s) \hspace*{-1mm}= \hspace*{-1mm} \int {(\lambda-s)^{-1}}{\rmp_\mJ (\lambda)} \dif \lambda$.

The decoupling property of \ac{rls} recovery indicates that, as $N$ grows large, $(x_n , \hx_n)$ converges in distribution to $(\xx_n,\hxx_n)$ when 
\begin{subequations}
\begin{align}
\sfp &= \lim_{N\uparrow \infty} \frac{1}{N} \sum_{n=1}^N \E{ \abs{\hxx_n-\xx_n}^2 } \\
\chi &= \lim_{N\uparrow \infty} \frac{1}{N} \sum_{n=1}^N \frac{\tau}{\theta^2} \E{ \left(\hxx_n-\xx_n\right) \zz^*}.
\end{align}
\end{subequations}
The proof follows the same approach as in \cite{bereyhi2018maximum} and is skipped due to similarity. For detailed derivations, the interested reader is referred to \cite{bereyhi2016rsb,bereyhi2016statistical}.

The decoupling principle can be intuitively interpreted as following: From each sample's point of view, the interference caused by other samples through linear mixture in measuring system is statistically equivalent to effective noise $\zz$, and the nonlinear joint recovery algorithm can be marginalized to a scalar \ac{rls} scheme by an effective tuning factor $\tau$. 
As a result of the decoupling principle, one can derive the asymptotic distortion in a closed form. 
Starting from \eqref{eq:dist}, we have
\begin{subequations}
\begin{align}
D &=  \lim_{N\uparrow \infty}  \frac{1}{N} \sum_{n=1}^N \E{ \sfd(\hx_n;x_n)}\\
&\stackrel{\dagger}{=}  \lim_{N\uparrow \infty} \frac{1}{N} \sum_{n=1}^N \E{ \sfd(\hxx_n;\xx_n)}  \label{eq:asymp_dist}
\end{align}
\end{subequations}
where $\dagger$ comes from the asymptotic decoupling property. In contrast to \eqref{eq:dist}, the expression in \eqref{eq:asymp_dist} is trivially calculated, since $\hxx_n$ is the output of a single-letter estimator.


\section{Algorithmic Approaches}
In this section, we intend to discuss the second question. To illustrate our objective, let us first discuss implementational perspectives of \ac{rls} recovery. 
\subsection{AMP Algorithms for Asymmetric RLS Recovery}
In practice, the computational complexity of \ac{rls} recovery depends on the algorithmic approach which solves the optimization problem in \eqref{eq:LSE}. For choices of the penalty function whose corresponding optimization problem is convex, generic linear programming can find the solution with feasible complexity. It is however known from the literature of \ac{cs} that \ac{rls} problems with large dimensions are most efficiently addressed via \ac{amp}. For a \ac{rls} problem, the \ac{amp} algorithm is derived by Gaussian approximations of loopy belief propagation algorithm \cite{donoho2009message,donoho2010message,rangan2011generalized,rangan2017vector}. Several generalized versions of \ac{amp}-based algorithms have been analyzed in the literature which enclose a large class of \ac{rls} recovery schemes and measuring matrices; see for example \cite{rangan2011generalized,rangan2017vector} and references therein. Considering the generic \ac{rls} recovery scheme in \eqref{eq:LSE} with the asymmetric penalty function \eqref{eq:penalty}, an \ac{amp}-based algorithm can be designed by adopting the generalized framework in \cite{rangan2011generalized} to this problem. As our focus in this section is to answer the tuning problem, we shorten the discussion by skipping the derivation of the algorithm and only discuss the approach. More details can be found in \cite{bereyhi2018precoding} where an \ac{amp} algorithm has been derived by a similar approach for a dual problem.

The \ac{rls} recovery algorithm can be considered as a max-sum problem in which the maxima of a marginalized posterior distribution is desired to be calculated. This is in fact the~Baye-sian interpretation of \ac{rls} recovery.  In \cite{rangan2011generalized}, the \ac{gamp} algorithm was proposed which develops an \ac{amp} algorithm for a max-sum problem of the form
\begin{align}
\bx = \argmax_{\bv\in\setX^N} \sum_{n=1}^N F(v_n,w_n)+\sum_{m=1}^M G(y_m,t_m) \label{eq:map}
\end{align}
for some given $F(\cdot,\cdot)$ and $G(\cdot,\cdot)$. In \eqref{eq:map}, the vector $\bt_{M\times 1}$ is a linear mapping of $\bv_{N\times 1}$, i.e., $\bt=\mA \bv$ for some matrix $\mA\in \setC^{M\times N}$. The entries of $\bv$ are generated from a known sequence $\set{w_n}$ via $\rmp(v_n|w_n)$ and the entries of $\by_{M\times 1}$ are given by passing the entries of $\bt_{M\times 1}$ through $\rmp(y_m|t_m)$. In the Bayesian framework, $F(\cdot,\cdot)$ and $G(\cdot,\cdot)$ model the postulated input and output distributions, and therefore, \eqref{eq:map} marginalizes the posterior for a given $\set{w_n}$. 

Considering the signal model in Section~\ref{sec:signal_model} and penalty in \eqref{eq:penalty}, the \ac{rls} recovery scheme in \eqref{eq:LSE} for $J=1$, is a max-sum problem with
\begin{subequations}
\begin{align}
F(v_n,w_n) = - u(v_n;w_n), \label{eq:f_in} \\
G(y_m,t_m) = - \abs{y_m-t_m}^2 \label{eq:f_out}
\end{align}
\end{subequations}
where the index $j$ is dropped since $J=1$. Consequently, the recovered samples $\bhx$ can be iteratively determined by adopting the \ac{gamp} algorithm to the max-sum problem in \eqref{eq:map} with these given priors; see \cite{bereyhi2018precoding} for more details. To extend the algorithm to multiple priors, i.e., $J >1$, one notes that the measuring system is rewritten as
\begin{align}
\by = \sum_{j=1}^J \mA_{\setN_j} \setB_j(\bx) + \bz
\end{align}
where $\mA_{\setN_j}\in \setC^{M \times \abs{\setN_j}} $ is constructed from the measuring matrix by collecting the columns indexed with $\setN_j$ and $\setB_j(\bx)$ denotes the $j$-th block of the signal samples. \ac{rls} recovery in this case solves the asymmetric max-sum problem 
\begin{align}
\bx = \argmax_{\bv\in\setX^N} \sum_{j=1}^J \sum_{n\in\setN_j} F_j(v_n,w_n)+\sum_{m=1}^M G(y_m,t_m) \label{eq:map2}
\end{align}
in which $\bv_{N\times 1}$ consists of $J$ blocks indexed by $\setN_j$ and the entries of block $j\in[J]$ are generated from their corresponding entries in $\set{w_n}$ with $n\in\setN_j$ via $\rmp_j(v_n|w_n)$. In this case as $N$ grows large, one can extends the analysis in \cite[Appendix C]{rangan2011generalized} to asymmetric versions by some lines of derivations. We however skip the details due to the page limitation.
\begin{remark}
The investigations in \cite{rangan2011generalized} consider \ac{iid} measuring matrices. For rotationally invariant matrices, as is considered in this paper, the framework has been recently generalized~in the literature; see for example \cite{rangan2017vector}. Consequently, one can straightforwardly extend the approach here to settings with non-\ac{iid} measuring matrices by invoking a generalized framework such as \ac{vamp}.
\end{remark}
\subsection{Tuning Strategy for Algorithmic Approaches}
The performance of \ac{rls} recovery depends on the tunable factors in the algorithm. Regarding the generic \ac{rls} scheme in \eqref{eq:LSE}, the performance is in general tuned via $\lambda$ as well as $\set{w_n}$ which need to be optimized in terms of the performance measure. Although \ac{amp}-based algorithms propose feasible approaches to address \ac{rls} recovery, the tuning problem remains unsolved and needs further investigations. For some particular \ac{rls} problems, optimal or suboptimal tuning strategies have been proposed through different analyses; see for example \cite{atitallah2017ber} and references therein. The tuning problem is however possible to be solved by the asymptotic results given via the replica method: \textit{Considering the decoupling property, the distortion in the large-system limit is derived as a function of $\lambda$ and empirical distribution of $\set{w_n}$. Consequently, $\lambda$ and $\set{w_n}$ are tuned such that the asymptotic distortion is minimized.} This minimization is feasible and can be addressed either numerically or analytically depending on the understudy setting. To be precise, we should indicate that this strategy is heuristic, since it assumes the two following postulates:
\begin{inparaenum}
\item[(1)] The large-system results given via the replica method~are accurate, and 
\item[(2)] There exists an integer $N_0$ such that for $N>N_0$, the tunable factors are constant in dimension $N$.
\end{inparaenum}
The validity of the former assumption is a sophisticated question in statistical mechanics and has been discussed for some particular cases. The second conjecture is however an interesting problem to investigate. For \ac{amp} frameworks studied in \cite{rangan2011generalized,rangan2017vector}, state evolution suggests that the latter assumption holds, since it shows that for a large class of settings, \ac{amp}-based algorithms asymptotically converge to the fixed-points derived via the replica method. The investigations in \cite{bereyhi2018precoding} moreover supports this assumption. In the sequel, we give further supports~by~con-sidering a particular example from the literature.
\section{Example of BPSK Recovery}
We consider the problem of BPSK recovery via regularized \ac{rls} with box and ordinary convex relaxations and discuss the optimal choice for the regularizer by employing our proposed tuning strategy. This problem was formerly studied in \cite{atitallah2017ber} where the error probability and the optimal~regularizer were found via the convex Gaussian min-max theorem \cite{thrampoulidis2015regularized}. 

In this problem, samples in $\bx$ are considered to be bipolar binaries, i.e., $x_n=\pm 1$, and the regularization term is considered to be $u(\bv)= \norm{\bv}^2/2$. In \ac{rls} recovery with box relaxation, the support is set to $\setX=[-1,1]$ which reduces \eqref{eq:LSE} to a convex problem. This relaxation in the ordinary convex case is $\setX=\setR$. The recovered samples are then calculated from $\bhx$ by $\bx^\star=\sgn{\bhx}$ where $\sgn\cdot$ is the sign function.

From the decoupling property, the asymptotic statistics of $(\hx_n, x_n)$ is characterized via the decoupled setup. Assuming $\mA$ to have zero-mean \ac{iid} entries with variance $1/M$, the decoupled system for the case with ordinary relaxation reads
\begin{align}
\hxx = (1+\tau)^{-1} {\yy}
\end{align}
where $\tau$ is given by
\begin{align}
\tau = \frac{\lambda +1-\rho + \sqrt{(\lambda +1 - \rho)^2 + 4 \lambda \rho} }{2\rho }
\end{align}
and $\yy=\xx + \zz$. Moreover, $\zz\sim\man(0,\theta^2 )$ where
\begin{align}
\theta^2 = \frac{\tau^2 + \sigma^2}{\rho (1+\tau)^2 -1}. \label{eq:theta1}
\end{align}
Under box relaxation, 
\begin{align}
\hxx = w((1+\tau)^{-1} {\yy})
\end{align}
where $w(x)=x$ for $\abs{x}\leq 1$ and $w(x)=\sgn x$ for $\abs{x}> 1$. 
Moreover, $\theta^2$ and $\tau$ satisfy 
\begin{subequations}
\begin{align}
2 \phi(\eta) + \frac{\xi}{\beta} \int_{-\eta}^\beta t^2 \md t+ \xi \phi(\beta) + \frac{\lambda}{\beta} &= \frac{\theta}{\rho} + \xi \phi(\eta) \label{eq:fix2-1} \\
\sigma^2 + 4  \rmQ(\eta) +  \frac{\xi^2}{\beta^2} \int_{-\eta}^\beta (t-\beta)^2 \md t &= \frac{\theta^2}{\rho} \label{eq:fix2-2}
\end{align}
\end{subequations}
for $\beta\coloneqq \theta^{-1}\tau$, $\xi\coloneqq (1+\tau)^{-1}\tau$ and $\eta\coloneqq \theta^{-1}(2+\tau)$.~Here, $\rmQ(\cdot)$ is the  $\rmQ$-function, $\phi(\cdot)$ denotes the real-valued zero-mean and unit variance Gaussian distribution and $\md t \coloneqq \phi(t) \dif t$. 

\begin{figure}[t]
\hspace*{-1cm}  
\resizebox{1.185\linewidth}{!}{
\pstool[width=.35\linewidth]{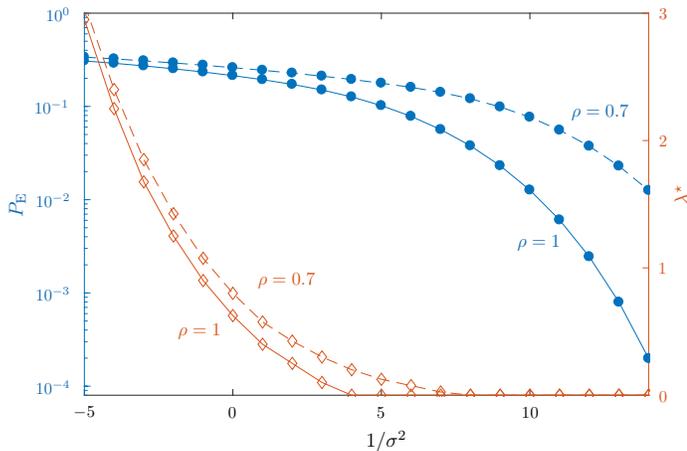}{
\psfrag{Pe}[c][t][0.24]{\textcolor{myblue}{$P_{\rm E}$}}
\psfrag{lam}[l][t][0.24]{\textcolor{myred}{$\lambda^\star$}}
\psfrag{SNR}[l][b][0.24]{$1/\sigma^2$}

\psfrag{r=07}[l][l][0.22]{\textcolor{myblue}{$\rho=0.7$}}
\psfrag{r=10}[l][l][0.22]{\textcolor{myblue}{$\rho=1$}}

\psfrag{r=07r}[l][l][0.22]{\textcolor{myred}{$\rho=0.7$}}
\psfrag{r=10r}[l][l][0.22]{\textcolor{myred}{$\rho=1$}}

\psfrag{x4}[r][c][0.2]{\textcolor{myblue}{$10^{-4}$}}
\psfrag{x3}[r][c][0.2]{\textcolor{myblue}{$10^{-3}$}}
\psfrag{x2}[r][c][0.2]{\textcolor{myblue}{$10^{-2}$}}
\psfrag{x1}[r][c][0.2]{\textcolor{myblue}{$10^{-1}$}}
\psfrag{x0}[r][c][0.2]{\textcolor{myblue}{$10^{0}$}}

%
\psfrag{3}[l][c][0.2]{\textcolor{myred}{$3$}}
\psfrag{2.5}[l][c][0.2]{\textcolor{myred}{$2.5$}}
\psfrag{2}[l][c][0.2]{\textcolor{myred}{$2$}}
\psfrag{1.5}[l][c][0.2]{\textcolor{myred}{$1.5$}}
\psfrag{1}[l][c][0.2]{\textcolor{myred}{$1$}}
\psfrag{0.5}[l][c][0.2]{\textcolor{myred}{$0.5$}}
\psfrag{0x}[c][c][0.2]{\textcolor{myred}{$0$}}

\psfrag{0}[c][c][0.2]{$0$}
\psfrag{-5}[c][c][0.2]{$-5$}
\psfrag{5}[c][c][0.2]{$5$}
\psfrag{10}[c][c][0.2]{$10$}

}}
\caption{Error probability and optimal tunable factor for regularized \ac{rls} recovery with box relaxation versus the \ac{snr}. Here, $\rho=M/N$.}
\label{fig:1}
\end{figure}

Considering either case, the asymptotic error probability $P_{\rm E}$ is determined by $\sfd(\hxx;\xx)=\mone \set{ \sgn{\hxx} \neq \xx}$ where $\mone\set\cdot$ is the indicator function. Substituting into \eqref{eq:asymp_dist}, we have
\begin{align}
P_{\rm E} = \rmQ \left( {\theta}^{-1} \right). \label{eq:P_E}
\end{align}
Comparing the results with \cite[Theorems 1 and 3]{atitallah2017ber}, one observes that the result for ordinary relaxation is derived from \eqref{eq:theta1}. For box relaxation, \cite[Theorems~3]{atitallah2017ber} represents $\theta$ as a solution to a min-max problem. The equivalency of this results to the fixed-point equations in \eqref{eq:fix2-1} and \eqref{eq:fix2-2} is numerically shown in Fig.~\ref{fig:1} where we have plotted the error probability, minimized over $\lambda$, versus $1/\sigma^{2}$ for two cases of $\rho=1$ and $\rho=0.7$. Invoking the asymptotic results, the optimal choice for $\lambda$ is found via our proposed tuning strategy which recovers the tuning results in \cite[Theorems 2 and 4]{atitallah2017ber} for box relaxation. 

\section{Conclusions}
This paper investigated an asymmetric form of \ac{rls} recovery. Using the replica method, we demonstrated the decoupling property of this setting and determined the asymptotic distortion for a generic distortion function. The results enabled us to propose a tuning strategy for algorithmic approaches to \ac{rls} recovery. As an example of our general framework, we recovered the available results for BPSK recovery with convex relations which was formerly studied in \cite{atitallah2017ber}. The systematic approach of this paper can address several other large \ac{rls} problems in the literature.




\bibliographystyle{IEEEtran}
\bibliography{refser}

\end{document}